\documentclass{appolb}
\usepackage{graphicx}
\usepackage{epsfig}

\begin{document}
\title{Exclusive heavy vector meson photoproduction on nuclei in NLO perturbative QCD%
\thanks{Presented at XXIX Cracow EPIPHANY Conference on Physics at the Electron-Ion Collider and Future Facilities, Cracow, Poland, 
January 16-19, 2023}%
}
\author{K.J. Eskola, V. Guzey, T. L\"oyt\"ainen, H. Paukkunen
\address{University of Jyvaskyla, Department of Physics, P.O. Box 35, FI-40014 University of Jyvaskyla, Finland and Helsinki Institute of Physics, P.O. Box 64, FI-00014 University of Helsinki, Finland}
\\[3mm]
{C.A. Flett
\address{Universit{\'e} Paris-Saclay, CNRS, IJCLab, 91405 Orsay, France}
}
}
\maketitle

\begin{abstract}

We make predictions for the cross section of coherent $J/\psi$ photoproduction in Pb-Pb and O-O 
ultraperipheral collisions (UPCs) at the LHC as a function of the $J/\psi$ rapidity $y$  in the framework of collinear factorization and next-to-leading order (NLO) perturbative QCD. We quantify the strong scale dependence and
significant uncertainties due to nuclear PDFs and show that our approach provides a reasonable description of the LHC data on coherent $J/\psi$ photoproduction in Pb-Pb UPCs. 
We demonstrate that these uncertainties are reduced by approximately a factor of 10 
in the scaled ratio of the O-O and Pb-Pb UPC cross sections. Our analysis indicates the dominance of the quark contribution to the UPC cross section at central rapidities, which affects the interpretation of the UPC data. 

\end{abstract}
  
\section{Introduction}
\label{sec:intro}

Studies of exclusive photoproduction of light and heavy vector mesons off protons and nuclei have been given a new impetus by
measurements of so-called ultraperipheral collisions (UPCs) at the Large Hadron Collider (LHC). 
In UPCs, ions (protons, nuclei) pass each other at large impact parameters, the short-range strong interaction between the
colliding hadrons is suppressed, and the reaction proceeds via emission of quasi-real photons, which are usually
treated in the Weizs\"acker-Williams equivalent photon approximation~\cite{Budnev:1975poe}.
At the LHC, the maximal energy of these photons reaches the TeV-range, which makes it effectively the highest energy photon collider to date.
Taking advantage of this, in UPCs these photons can be used as a probe to study open questions of the proton and nucleus structure 
and the strong interaction dynamics in 
quantum chromodynamics (QCD) as well as to search for new physics~\cite{Bertulani:2005ru,Baltz:2007kq,Contreras:2015dqa,Klein:2019qfb}. 

One of the most thoroughly studied UPC processes at the LHC is exclusive photoproduction of heavy vector mesons, in particular,
photoproduction of $J/\psi$. The interest in it is driven by the original observation that in the leading double logarithmic approximation
of perturbative QCD (pQCD),
the $J/\psi$ photoproduction cross section is directly proportional to the small-$x$ gluon density of the target squared~\cite{Ryskin:1992ui}. 

The application of this idea to nuclear targets allows for a direct comparison of the nuclear suppression factor extracted from the data on coherent $J/\psi$ photoproduction in lead-lead (Pb-Pb) UPCs at the LHC with small-$x$ nuclear modifications of 
the nuclear gluon density~\cite{Guzey:2013xba,Guzey:2013qza,Guzey:2020ntc}.
The good agreement with the ALICE data at the $J/\psi$ rapidity $y \approx 0$ 
gives direct evidence of the significant $\sim 40$\% gluon nuclear shadowing (suppression) in lead
at $x \approx 6 \times 10^{-4}-10^{-3}$, 
in agreement with the predictions of the leading twist approach~\cite{Frankfurt:2011cs} and the EPPS16 nuclear parton distribution function (nPDFs)~\cite{Eskola:2016oht}.

Further progress in obtaining new constraints on proton and nucleus 
parton distribution functions (PDFs) using $J/\psi$ photoproducton in UPCs requires both experimental and theoretical efforts.
One perspective direction is studies of UPCs accompanied by forward neutron emission~\cite{Guzey:2013jaa}, which allows one to separate the high-$W$ and low-$W$ contributions to the UPC cross section at a given $y$ ($W$ is the photon-nucleon invariant energy) and, hence, to probe much lower values of $x$ down to $x \approx 6 \times 10^{-5}$~\cite{CMS:2022nnw}.
On the theory side, significant advances have been made to refine calculations of exclusive $J/\psi$ photoproduction. 
These include an analysis of effects associated with the gluon and charm quark transverse momenta in the high-energy factorization
approach~\cite{Ryskin:1995hz}, the calculation of next-to-leading order (NLO) pQCD corrections in the framework of collinear factorization~\cite{Ivanov:2004vd,Jones:2015nna} and their 
taming using the $Q_0$-subtraction method~\cite{Flett:2019pux}, and estimates of relativistic corrections~\cite{Lappi:2020ufv}
and pre-asymptotic effects~\cite{Frankfurt:1997fj} associated with the charmonium light-cone wave function.

In this paper, we summarize our predictions for coherent $J/\psi$ photoproduction in Pb-Pb and oxygen-oxygen (O-O) UPCs at the LHC in the 
framework of collinear factorization and NLO pQCD~\cite{Eskola:2022vpi,Eskola:2022vaf}. We quantify the strong scale dependence and
significant PDF-related uncertainties of our predictions.  We show that NLO pQCD provides a reasonable description of the 
LHC data on coherent $J/\psi$ photoproduction in Pb-Pb UPCs, and also that the uncertainties are reduced by approximately a factor of 10 in the ratio of the O-O and Pb-Pb UPC cross sections.
A surprising consequence of our analysis is the dominance of the quark 
contribution at central rapidities, which challenges the interpretation of the data in terms of the small-$x$ nuclear gluon distribution.

\section{Exclusive $J/\psi$ photoproduction on nuclei in NLO pQCD}
\label{sec:nlo}

The cross section of exclusive coherent $J/\psi$ photoproduction in Pb-Pb UPCs is given by a sum
of two terms since both of the colliding ions can serve as a photon source and a target~\cite{Baltz:2007kq},
\begin{equation}
\frac{d\sigma}{dy}=\left[k \frac{dN_{\gamma/A}}{dk} \sigma^{\gamma A \to J/\psi A}\right]_{k=k^{+}}+
\left[k \frac{dN_{\gamma/A}}{dk} \sigma^{\gamma A \to J/\psi A}\right]_{k=k^{-}} \,,
\label{eq:upc}
\end{equation}
where $y$ is the $J/\psi$ rapidity and $k dN_{\gamma/A}/dk$ is the photon flux calculated in the equivalent-photon
approximation, 
see details in~\cite{Eskola:2022vpi}.
Therefore, for a given value of $y$, 
there is a two-fold ambiguity in the photon energy, $k^{\pm}=(M_{J/\psi}/2)e^{\pm y}$, where $M_{J/\psi}$ is the $J/\psi$
mass.

The information on the nuclear partonic structure and the strong interaction dynamics is contained in the underlying
$\sigma^{\gamma A \to J/\psi A}$ cross section of $J/\psi$ photoproduction.
Assuming that the charm quark mass $m_c$ provides a hard scale so that collinear factorization for hard exclusive processes is applicable~\cite{Collins:1996fb}, the $\gamma+A \to J/\psi+A$
amplitude to NLO accuracy is given by the following convolution~\cite{Ivanov:2004vd,Jones:2015nna},
\begin{equation}
{\cal M} \propto \sqrt{\langle O_1 \rangle_{V}} \int^{1}_{-1}dx \left[T_g(x,\xi)F_A^g(x,\xi,t,\mu)
+T_q(x,\xi)F_A^{q,S}(x,\xi,t,\mu)\right] \,,
\label{eq:amp}
\end{equation}
where $\langle O_1 \rangle_{V}$ is the non-relativistic QCD (NRQCD) matrix element determined from the $J/\psi \to l^{+} l^{-}$
leptonic decay, $T_g(x,\xi)$ and $T_q(x,\xi)$ are the gluon and quark NLO coefficient functions, and 
$F_A^g(x,\xi,t,\mu)$ and 
\allowbreak
$F_A^{q,S}(x,\xi,t,\mu)$ are the gluon and quark singlet nuclear generalized parton distribution functuons (GPDs),
respectively. The GPDs depend on the light-cone momentum fractions $x$ and $\xi$, the momentum transfer to the target squared $t$,
and the scale $\mu$. 
In our analysis, we take the renormalization and factorization scales to be equal and allow them to vary in the 
interval $m_c < \mu < 2 m_c$.
The derivations in~\cite{Ivanov:2004vd,Jones:2015nna} assume the static approximation for the charmonium wave function~\cite{Hoodbhoy:1996zg}, whose consistency requires that $m_c=M_{J/\psi}/2$. 

\begin{figure}[htb]
\centerline{%
\includegraphics[width=6cm]{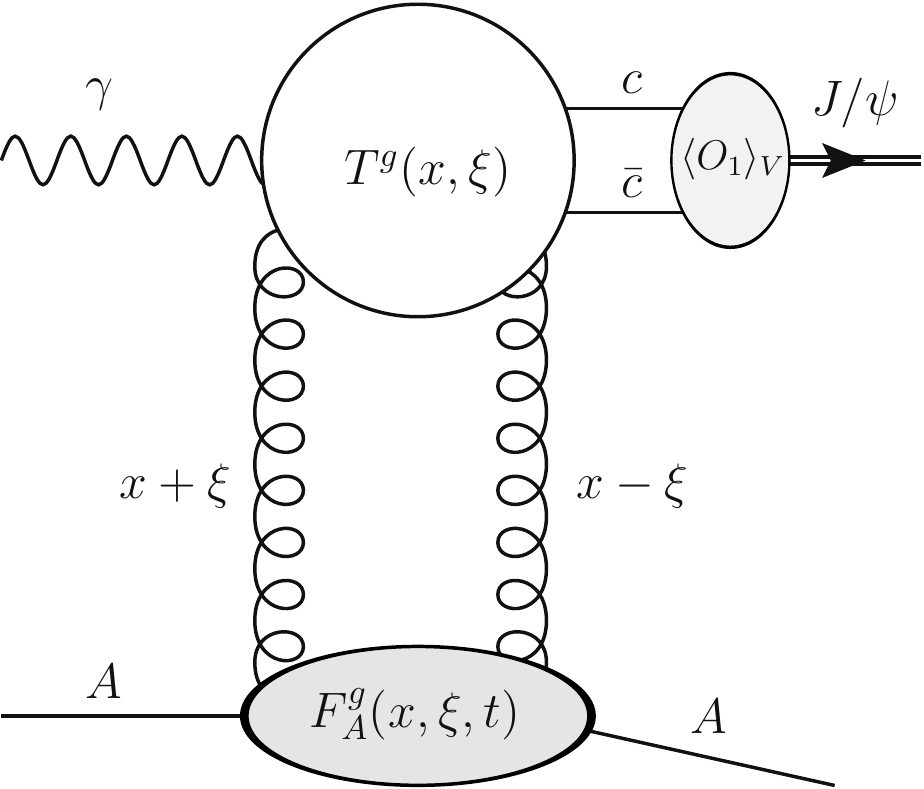}
\hspace{0.75cm}
\includegraphics[width=6cm]{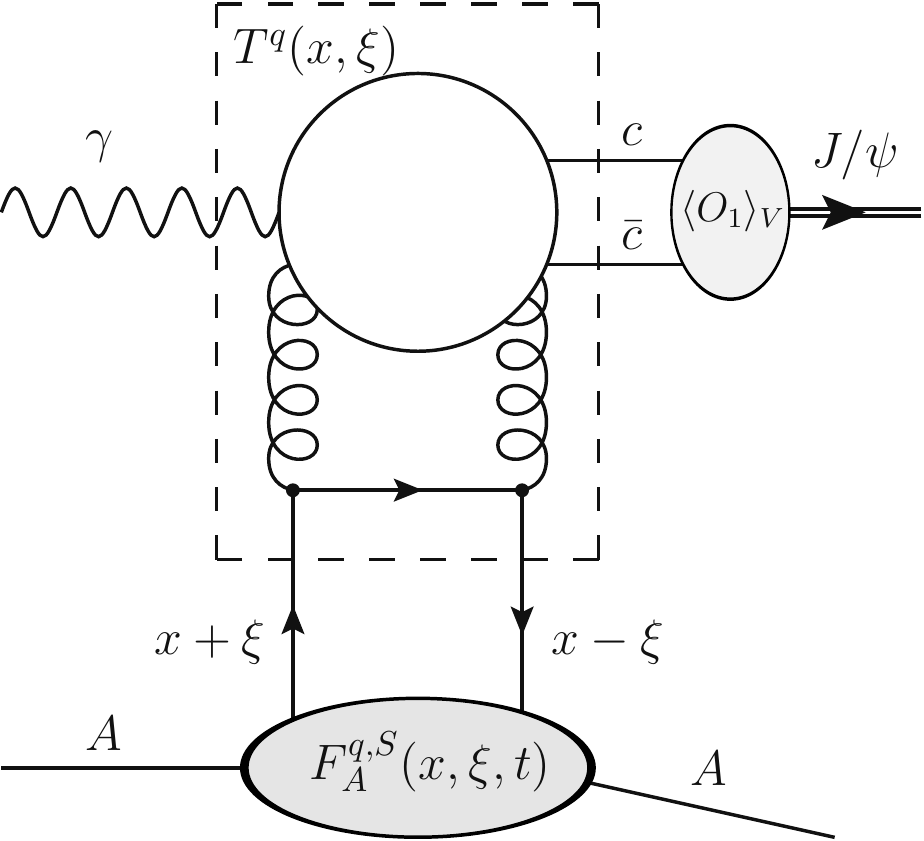}}
\caption{Graphical representation of the $\gamma+A \to J/\psi+A$ amplitude in NLO pQCD: gluon (left) and quark (right) contributions.}
\label{fig:amp}
\end{figure}

In the graphical form, the amplitude of Eq.~(\ref{eq:amp}) is presented in Fig.~\ref{fig:amp} showing gluon (left)
and quark (right) contributions. At leading order (LO), there is only the gluon term, while at NLO both the gluons and quarks 
contribute.

The parameter $\xi$ in Eq.~(\ref{eq:amp}) quantifies the imbalance (so-called skewness) of the momentum fractions carried by the partons attached to the target (see Fig.~\ref{fig:amp}) and is fixed by the process kinematics. In the limit of high $W$, 
$\xi \approx M_{J/\psi}^2/(2W^2) \ll 1$, which in general simplifies the modeling of GPDs. Indeed, for sufficiently 
large values of $\mu$, the $\mu^2$ evolution of GPDs almost completely washes out the information on the $\xi$-dependence of the GPDs at 
some low starting scale $\mu_0 \sim 1$ GeV~\cite{Dutrieux:2023qnz}, where they can hence be taken in the $\xi \to 0$ limit. 
As a result, one can relate GPDs to PDFs~\cite{Shuvaev:1999ce}.

In our work, we neglect the $\xi$-dependence of the nuclear GPDs because its effects are significantly smaller than
the theoretical uncertainties associated with the choice of the scale of $\mu$ and those due to nuclear PDFs, which we
consider in detail in Sec.~\ref{sec:results}. Employing the symmetries of GPDs, we use
\begin{eqnarray}
F_A^g(\pm x,\xi,t,\mu) &=&xg_A(x,\mu) F_A(t) \,, \nonumber\\
F_A^{q,S}(\pm x,\xi,t,\mu) &= & \sum_{q=u,d,s,c}\left[\theta(x)q_A(x,\mu)-\theta(-x){\bar q}_A(x,\mu)\right] F_A(t) \,, 
\end{eqnarray}
where $x \in [0,1]$, $F_A(t)$ is the nuclear form factor, $g_A(x,\mu)$, $q_A(x,\mu)$ and ${\bar q}_A(x,\mu)$ are nuclear gluon, quark and antiquark PDFs.
We use the Fermi model (Woods-Saxon) for the nuclear density~\cite{DeVries:1987atn} in the calculation of
the nuclear form factor and such state-of-the-art nuclear PDFs as EPPS16~\cite{Eskola:2016oht}, 
nNNPDF2.0~\cite{AbdulKhalek:2020yuc}, EPPS21~\cite{Eskola:2021nhw}, nNNPDF3.0~\cite{AbdulKhalek:2022fyi}, and
nCTEQ15WZSIH~\cite{Duwentaster:2021ioo}.

\section{Predictions for coherent $J/\psi$ photoproduction in Pb-Pb and O-O UPCs at the LHC}
\label{sec:results}

The formalism presented in the previous section allows us to make predictions for the cross sections of coherent $J/\psi$ photoproduction in Pb-Pb and O-O UPCs at the LHC as a function of the $J/\psi$ rapidity $y$.

\begin{figure}[tb]
\centerline{%
\includegraphics[width=11.cm]{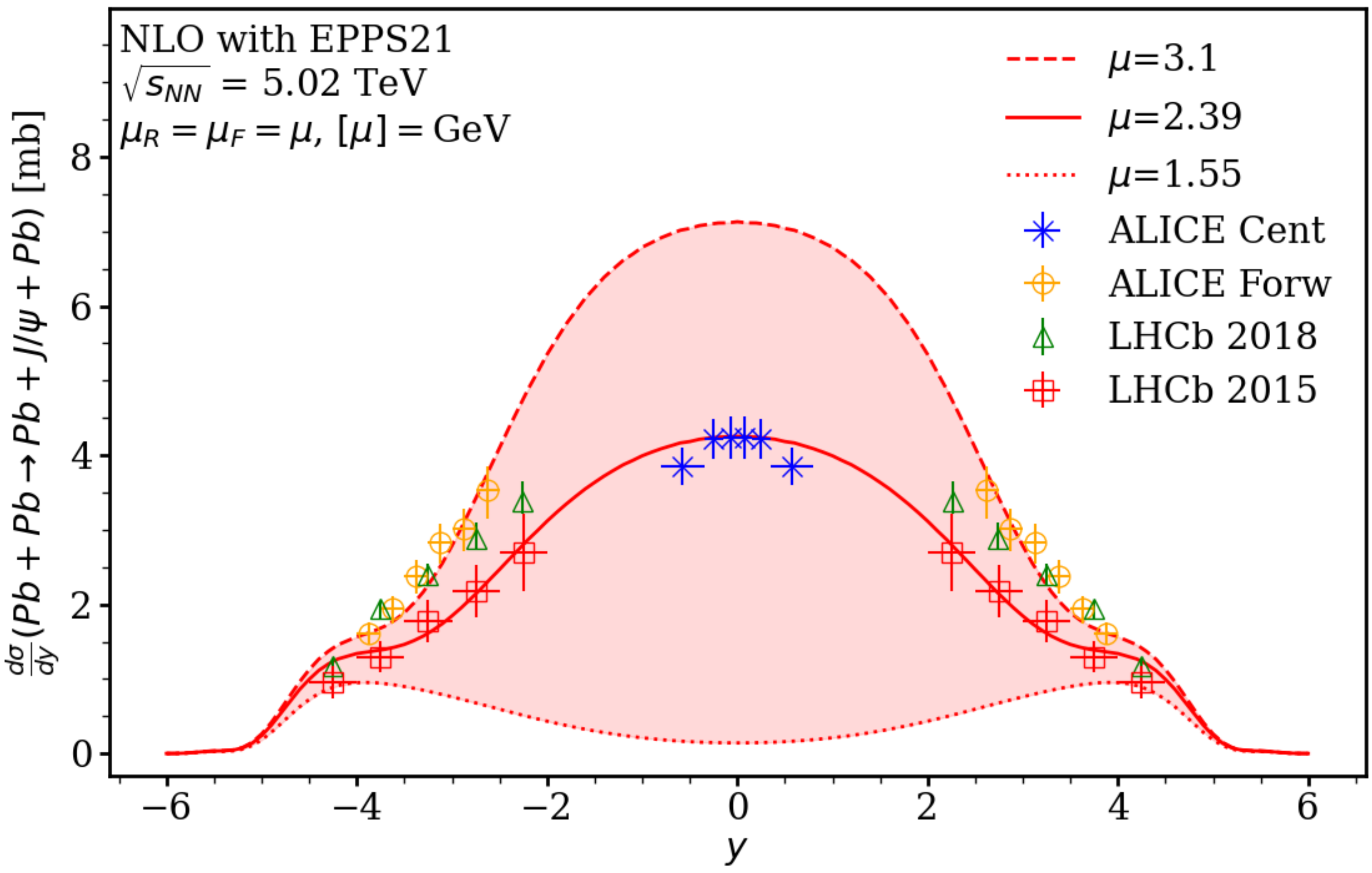}}
\caption{NLO pQCD predictions for the cross section of coherent $J/\psi$ photoproduction
in Pb-Pb UPCs at $\sqrt{s_{NN}}=5.02$ TeV as a function of the $J/\psi$ rapidity $y$
 and their comparison with the LHC Run 2 data. 
The calculations correspond to the EPPS21 nPDFs and $\mu=M_{J/\psi}/2=1.55$ GeV (lower boundary), $\mu=M_{J/\psi}=3.1$ GeV (upper boundary), and $\mu=2.39$ GeV (optimal scale). From~\cite{Eskola:2022vaf}.}
\label{fig:EPPS21_run2_scale}
\end{figure}

Figure~\ref{fig:EPPS21_run2_scale} quantifies the scale dependence of our NLO pQCD predictions for 
$d\sigma^{{\rm Pb}+{\rm Pb} \to {\rm Pb}+J/\psi+{\rm Pb}}/dy$ at $\sqrt{s_{NN}}=5.02$ TeV and compares them with all available Run 2 data, 
see~\cite{Eskola:2022vpi} for references. The calculations are carried out with the central EPPS21 nPDFs 
and at the values of the scale $\mu$ spanning the $M_{J/\psi}/2 < \mu < M_{J/\psi}$ range.  One can see from the figure that while the scale dependence is very strong, 
 it is possible to find an optimal scale, $\mu=2.39$ GeV in the considered case, which provides a 
 reasonable (good at $y \sim 0$ and worse at large $|y|$) description   of the LHC data on coherent $J/\psi$ photoproduction in Pb-Pb UPCs, so
 that the data at all $y$ lie within our scale uncertainty envelope.

 Note that while the use of other nuclear PDFs leads to similar results at central rapidities, there are certain differences 
 at large $|y|$. In particular, the nCTEQ15WZSIH nPDFs provide a somewhat better agreement with the forward ALICE and 2018 LHCb data, 
 which may hint on the sensitivity to the strange quark distributions in nuclei.
 
\begin{figure}[tb]
\centerline{%
\includegraphics[width=11.cm]{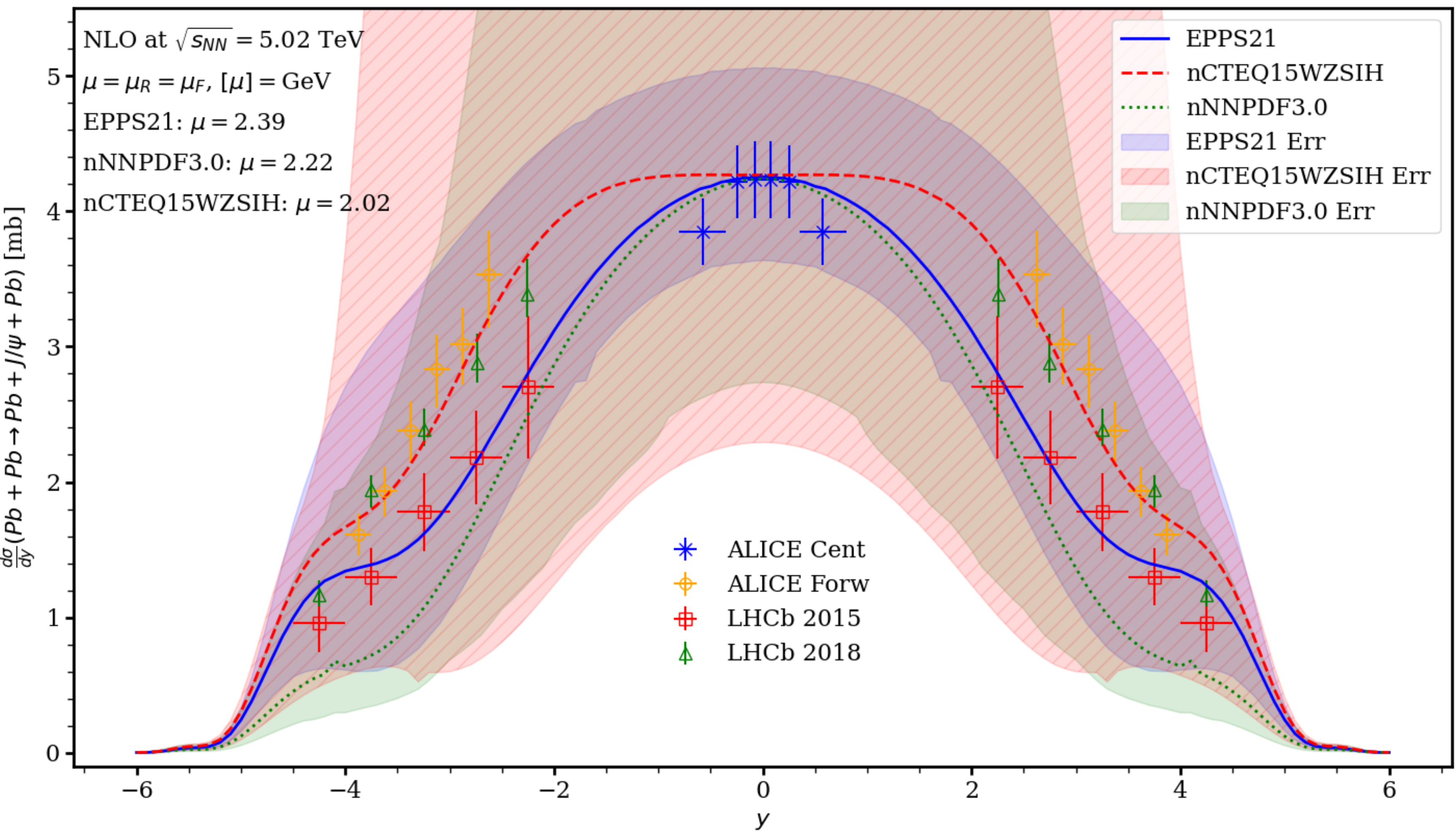}}
\caption{NLO pQCD predictions for 
$d\sigma^{{\rm Pb}+{\rm Pb} \to {\rm Pb}+J/\psi+{\rm Pb}}/dy$ at 5.02 TeV
as a function of the $J/\psi$ rapidity $y$ for EPPS21, nNNPDF3.0, and nCTEQ15WZSIH nPDFs, see text for details. For comparison, the Run 2 LHC data are also shown.  From~\cite{Eskola:2022vaf}.}
\label{fig:EPPS21_run2_error}
\end{figure}

In Fig.~\ref{fig:EPPS21_run2_error}, we show our predictions for $d\sigma^{{\rm Pb}+{\rm Pb} \to {\rm Pb}+J/\psi+{\rm Pb}}/dy$ at $\sqrt{s_{NN}}=5.02$ TeV using different nPDFs. The curves correspond to
the central EPPS21 (blue solid), nCTEQ15WZSIH (red dashed), 
and nNNPDF3.0 (green dotted) nPDFs along with their propagated uncertainties given by the respective shaded bands.
The calculations are performed at the optimal scales corresponding to each PDF set; the quality of the agreement with the data is illustrated by a comparison with the Run 2 data. One can see from the figure that the uncertainties due to nuclear PDFs are quite significant and exceed the experimental errors. Hence, this can be viewed as an opportunity to improve the determination of nuclear PDFs using the data 
on $J/\psi$ photoproduction in nucleus-nucleus UPCs.

Note that compared to our results in~\cite{Eskola:2022vpi}, the abnormally large uncertainty associated with the EPPS16 
nPDFs (in fact with the CT14 baseline PDFs) disappears, when using the more recent EPPS21 nPDFs. Also, since the nNNPDF3.0 nPDFs correspond to a much less constrained fit, the corresponding nuclear PDFs uncertainties are the largest among those shown in Fig.~\ref{fig:EPPS21_run2_error}.
Note also that the nCTEQ15WZSIH errors do not include the free proton PDF uncertainties.

\begin{figure}[tb]
\centerline{%
\includegraphics[width=12.5cm]{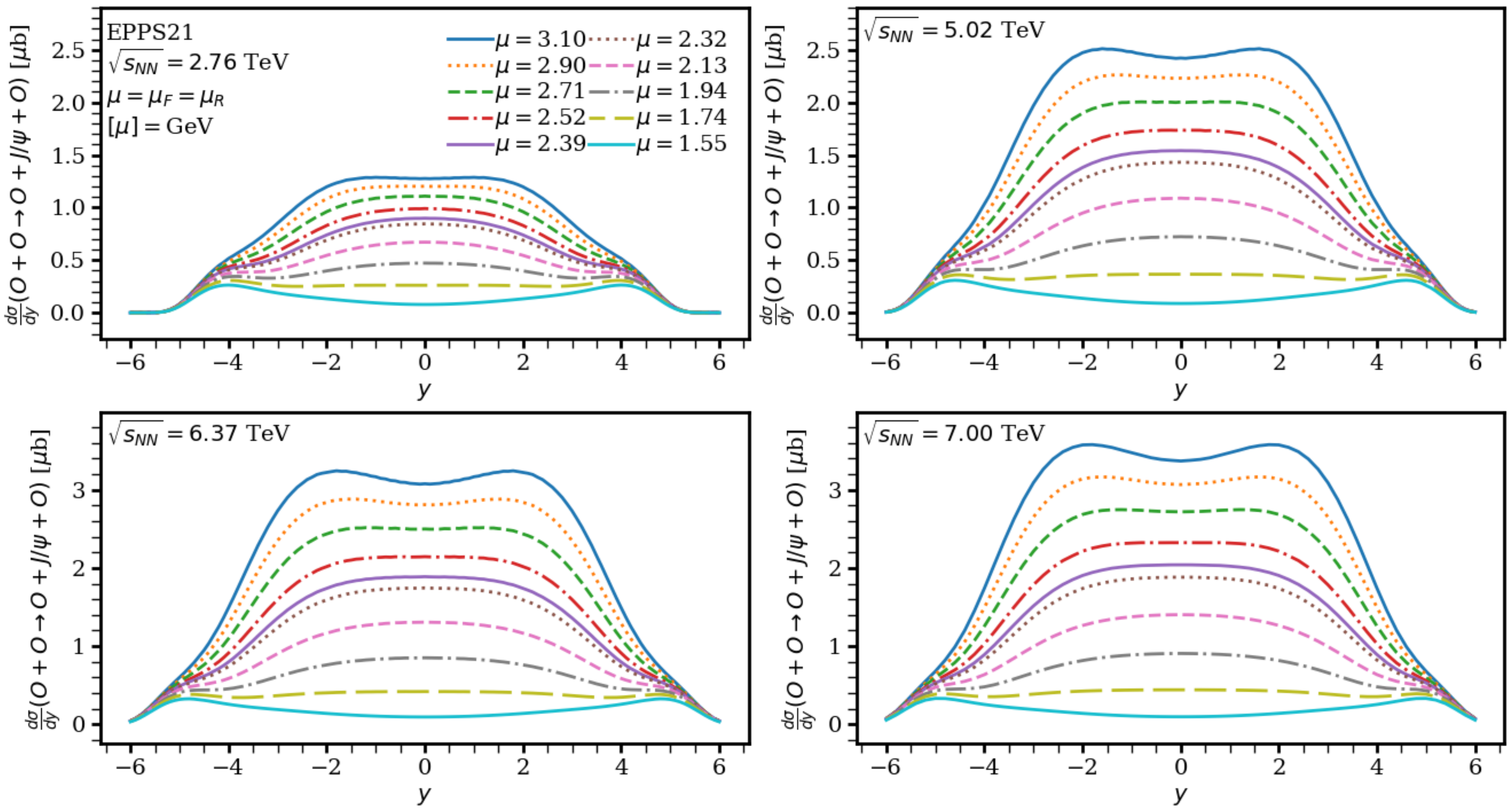}}
\caption{NLO pQCD predictions for the cross section of coherent $J/\psi$ photoproduction in O-O UPCs at $\sqrt{s_{NN}}=2.76$, 5.02, 6.37 and 7 TeV as a function of the $J/\psi$ rapidity $y$.
 The curves correspond to the predictions based on central EPPS21 nPDFs and several values of 
 $\mu$ in the interval of
$m_c < \mu < 2 m_c$. From~\cite{Eskola:2022vaf}.
}
\label{fig:oxygen}
\end{figure}

Anticipating the oxygen-oxygen (O-O) run at the LHC, we also made predictions for the cross section of coherent $J/\psi$ photoproduction in O-O UPCs for four collision energies, $\sqrt{s_{NN}}=2.76$, 5.02, 6.37 and 7 TeV. Our results for 
$d\sigma^{{\rm O}+{\rm O} \to {\rm O}+J/\psi+{\rm O}}/dy$ as a function of $y$ are shown in Fig.~\ref{fig:oxygen}. The calculations correspond 
to using the central EPPS21 nPDFs as input and a range of scales 
$m_c < \mu < 2 m_c$. 
As in the case of Pb-Pb UPCs, we find that the scale dependence and the uncertainty due to nPDFs are large.

\begin{figure}[tb]
\centerline{%
\includegraphics[width=12.5cm]{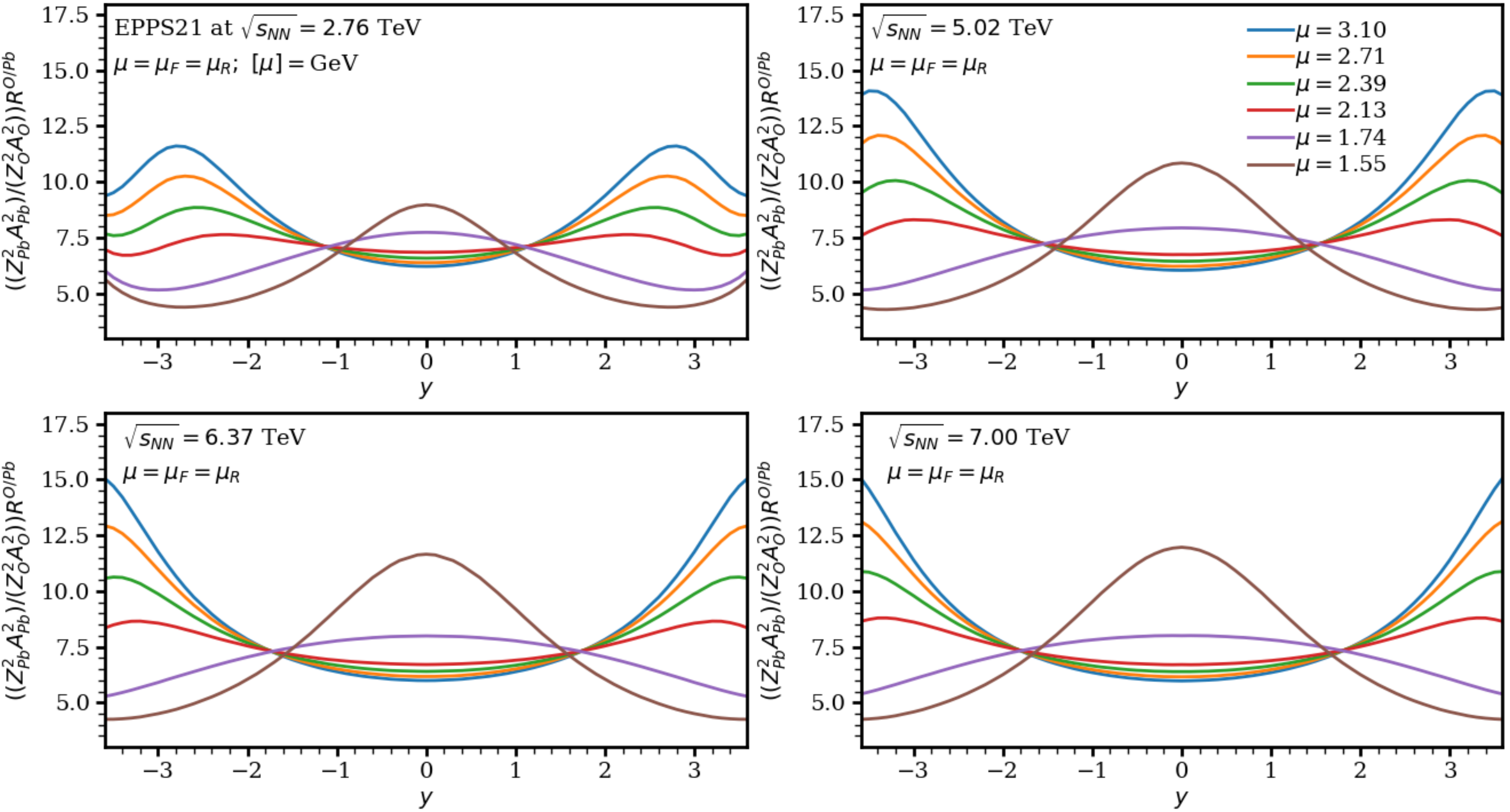}}
\caption{The scaled ratio of the O-O and Pb-Pb UPC cross sections, see Eq.~(\ref{eq:R}), as a function of the
$J/\psi$ rapidity $y$ for four values of $\sqrt{s_{NN}}$. See Fig.~\ref{fig:oxygen} for the labeling of the curves.
From~\cite{Eskola:2022vaf}.
}
\label{fig:oxygen_ratio}
\end{figure}

One can reduce these uncertainties by considering the following scaled ratio of the oxygen-oxygen to lead-lead UPC cross sections, 
\begin{eqnarray}
\tilde{R}^{{\rm O}/{\rm Pb}} &=& \left(\frac{208 Z_{\rm Pb}}{16 Z_{\rm O}} \right)^2 \frac{d\sigma({\rm O}+{\rm O} \to {\rm O}+J/\psi+{\rm O})/dy}{d\sigma({\rm Pb}+{\rm Pb} \to {\rm Pb}+J/\psi+{\rm Pb})/dy} \nonumber\\
&=& \left(\frac{208 Z_{\rm Pb}}{16 Z_{\rm O}} \right)^2 R^{{\rm O}/{\rm Pb}}\,,
\label{eq:R}
\end{eqnarray}
where $Z_{\rm Pb}=79$ and $Z_{\rm O}=8$ are the nucleus electric charges. The ratios $\tilde{R}^{{\rm O}/{\rm Pb}}$ as a function of $y$ for $\sqrt{s_{NN}}=2.76$, 5.02, 6.37 and 7 TeV are shown in Fig.~\ref{fig:oxygen_ratio}. 
For the labeling of the curves, see Fig.~\ref{fig:oxygen}.
A comparison with the results in 
Figs.~\ref{fig:EPPS21_run2_scale} and \ref{fig:oxygen}
 demonstrates that in the ratio of the cross sections for different nuclei, 
 the scale dependence is reduced by approximately a factor of 10 compared to the individual Pb-Pb and O-O cross sections.
This can be explained by the fact that since the NLO coefficient functions for the oxygen and lead targets are the same, the 
differences come mostly from the oxygen and lead nPDFs. Additionally, there is a  partial cancellation of uncertainties associated with proton PDFs.

\begin{figure}[htb]
\centerline{%
\includegraphics[width=12.5cm]{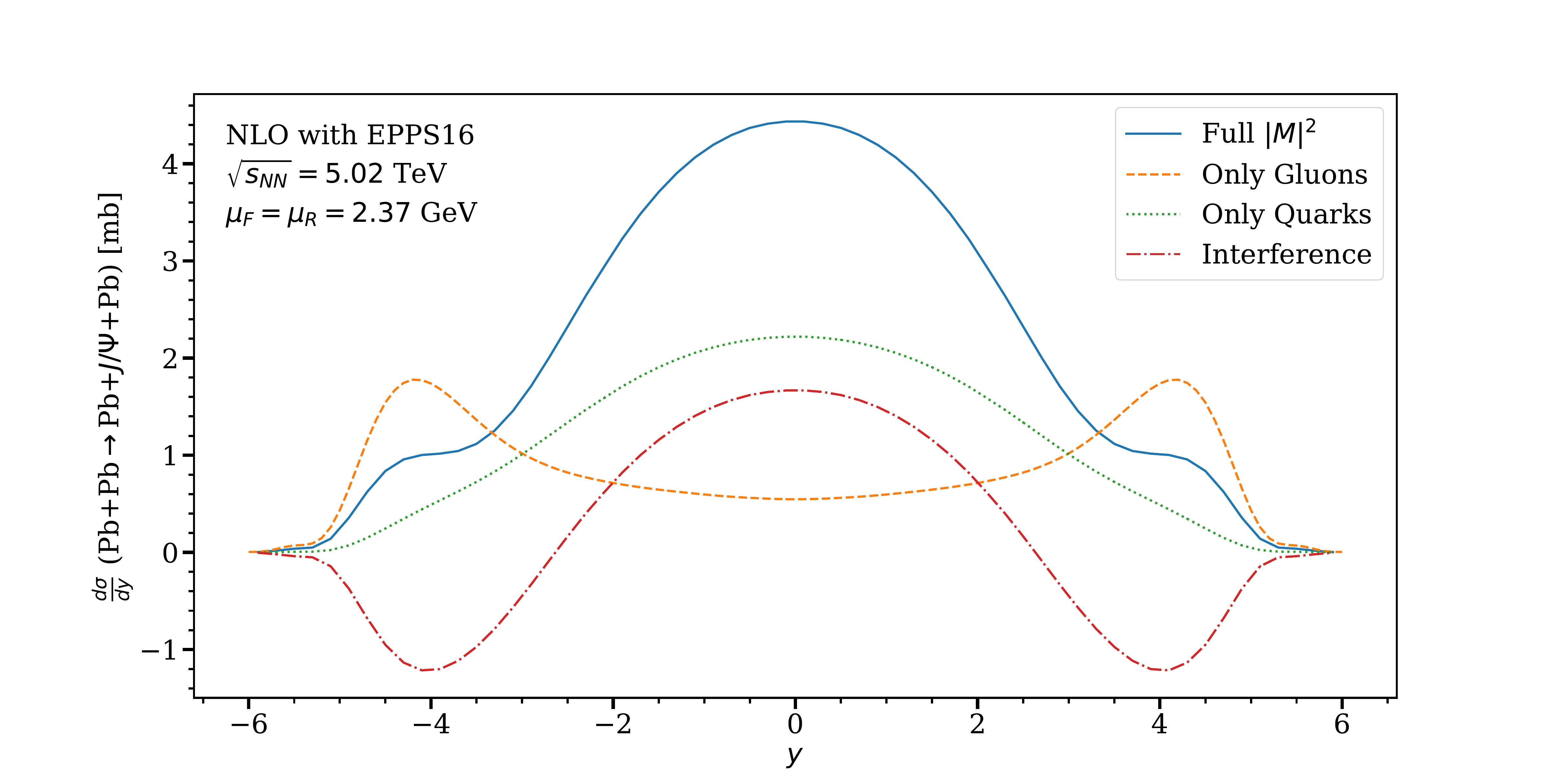}}
\caption{The separate quark, gluon and their interference contributions to
$d\sigma^{{\rm Pb}+{\rm Pb} \to {\rm Pb}+J/\psi+{\rm Pb}}/dy$ as a function of $y$ at $\sqrt{s_{NN}}=5.02$ TeV.
From~\cite{Eskola:2022vpi}.}
\label{fig:quark}
\end{figure}

One of the most striking conclusions of our analyses is the observation that the quark contribution dominates the 
cross section of $J/\psi$ photoproduction in Pb-Pb and O-O UPCs at non-forward rapidities, which originates from
the strong cancellations between LO and NLO gluon terms in Eq.~(\ref{eq:amp}). This is illustrated in 
Fig.~\ref{fig:quark} showing the separate quark, gluon and their interference contributions to
$d\sigma^{{\rm Pb}+{\rm Pb} \to {\rm Pb}+J/\psi+{\rm Pb}}/dy$ as a function of $y$ at $\sqrt{s_{NN}}=5.02$ TeV.
The curves correspond to the EPPS16 nPDFs and $\mu=2.37$ GeV is the 
optimal scale.
One can see from the figure that the quarks provide the dominant contribution for $|y| < 2$.

At face value, this changes the interpretation of the data on coherent $J/\psi$ photoproduction in heavy-ion UPCs as a probe of small-$x$ nuclear gluons and replaces it with an observation that both the magnitude and shape of the rapidity dependence
of the UPC cross section depend on an interplay 
of
the quark and gluon contributions. While this is most likely a feature of NLO pQCD, we anticipate that at 
next-to-next-leading order (NNLO), mixing of quark terms at both NLO and NNLO should also play a role.

\section{Conclusions}

Using the framework of collinear factorization and NLO pQCD, we made predictions for the rapidity dependence of the cross section 
of coherent $J/\psi$ photoproduction in Pb-Pb and O-O UPCs at the LHC. We quantified the strong scale dependence and
significant uncertainty due to nuclear PDFs of our predictions and showed that they provide a reasonable description of the 
LHC data on coherent $J/\psi$ photoproduction in Pb-Pb UPCs. We demonstrated that these uncertainties are dramatically reduced 
in the scaled ratio of the O-O and Pb-Pb UPC cross sections. Also, our analysis indicated the dominance of the quark contribution to the UPC cross section at central rapidities, so the interpretation of the UPC data must be taken with care.

\section*{Acknowledgements}
We acknowledge the financial support from the Magnus Ehrnrooth foundation (T.L.), the Academy of Finland
projects 308301 (H.P.) and 330448 (K.J.E.). This research was funded as part of the Center of Excellence in
Quark Matter of the Academy of Finland (projects 346325 and 346326)
and of the European
Research Council project ERC-2018-ADG-835105 YoctoLHC.

\end{document}